**Intégration dans l'économie-monde et régionalisation de l'espace centre-est européen depuis 1989**

Natalia Zdanowska

Centre for Advanced Spatial Analysis, University College London,
UMR 8504 Géographie-cités, Université Paris 1 Panthéon-Sorbonne
n.zdanowska@ucl.ac.uk

**Résumé**

La chute du Mur de Berlin en 1989, a modifié les relations unissant les villes de l'ancien bloc communiste. La réorientation européenne et mondiale des interactions qui en a découlé soulève, aujourd'hui, la question du devenir des relations historiques entre les villes au sein même de l'espace centre-est européen, mais également avec celles de l'ex-URSS et de l'ex-Yougoslavie. Peut-on dire que les villes centre-est européennes reproduisent, dans un contexte économique nouveau, des trajectoires déjà empruntées par le passé ?

Cet article s'efforcera de retracer leurs évolutions dans le cadre des échanges commerciaux et des connexions aériennes depuis 1989. Il se propose d'identifier ces traces du passé au sein des réseaux de firmes transnationales pour les années plus récentes. La principale contribution de ce travail est de montrer qu'il y a eu formation progressive de plusieurs régions économiques en Europe centrale et orientale à la suite de l'intégration dans l'économie-monde.

**Mots-clés** : Europe centrale et orientale, villes, réseaux commerciaux, réseaux de trafic aérien, réseaux de firmes, relations du passé, régionalisation

**Integration into *économie-monde* and regionalisation of the Central Eastern European space since 1989**

**Abstract**

The fall of the Berlin Wall in 1989, modified the relations between cities of the former communist bloc. The European and worldwide reorientation of interactions that followed raises the question of the actual state of historical relationships between Central Eastern European cities, but also with ex-USSR and ex-Yugoslavian ones. Do Central and Eastern European cities reproduce trajectories from the past in a new economic context?

This paper will examine their evolution in terms of trade exchanges and air traffic connexions since 1989. They are confronted with transnational firm networks for the recent years. The main contribution is to show a progressive formation of several economic regions in Central and Eastern Europe as a result of integration into Braudel's *économie-monde*.

**Key-words**: Central and Eastern Europe, cities, trade networks, air traffic networks, firm networks, relations from the past, regionalisation

**Introduction**

L'ampleur et la rapidité des transformations structurelles qu'a connues l'Europe centrale et orientale[1] après 1989 semblent avoir « pris en défaut la 'soviétologie' occidentale » (Coudroy de Lille, 2016, p. 2). Même Richard Pipes, l'un des plus grands historiens américains de l'URSS et de la Russie, n'a pas su prévoir la chute du communisme (Werth, 1999). Cet étonnement est à l'image du bouleversement économique, politique et sociétal qui en est résulté dans l'ensemble de l'ex-bloc de l'Est.

La chute du Mur de Berlin a bien évidemment étendu aux pays d'Europe centrale et orientale l'influence des processus d'européanisation – ou d'intégration à l'Union européenne (Bafoil, 2006) –, et de mondialisation, déjà en cours à l'ouest de l'Europe. Toutefois ce changement ne peut être apprécié sans prendre en considération les singularités des trajectoires de chacun de ces pays. En effet, l'histoire récente, ou plus ancienne, révèle la mise en place de relations privilégiées entre certaines villes centre-est européennes. L'enjeu est de savoir dans quelle mesure ces relations passées ont perduré, compte tenu de la mise en place depuis 1989 de logiques de domination exercées par certaines villes occidentales remontant à la formation de l'*économie-monde* (Braudel, 1979). Ce terme est entendu comme « morceau de la planète économiquement autonome, capable pour l'essentiel de se suffire à lui-même et auquel ses liaisons et ses échanges intérieurs confèrent une certaine unité organique » (Braudel, 1979, p. 20). Il traduit un phénomène beaucoup plus fort que le seul terme de *mondialisation*, désignant le produit de l'ensemble des diffusions, des échanges et communications entre différentes parties du monde à l'échelle géographique la plus haute, le « mondial » (Dollfus et al., 1999).

Cet article se penchera sur l'étude de l'intégration des villes centre-est européenne dans *l'économie-monde* par les interactions. Les *interactions* entre villes sont en géographie humaine l'expression de relations entre les lieux et des influences qu'ils exercent l'un sur l'autre (Sanders, 2001 ; Grasland, 2004). L'*intégration* met en relation la forme et l'intensité des interactions entre les entités d'un système (Sanders, 2012). Le degré d'intégration d'un sous-ensemble est d'autant plus élevé que les relations entre les parties constitutives du sous-ensemble apparaissent systématiquement plus intenses à l'intérieur qu'en direction d'unités spatiales extérieures (Saint-Julien, 2004).

Nous supposons que malgré les logiques européennes et mondiales prédominantes, les relations au sein même de l'Europe centrale et orientale établies avant 1989 perdurent jusqu'à aujourd'hui. Ceci relèverait alors du phénomène de résilience, ou « capacité d'un système à pouvoir intégrer dans son fonctionnement une perturbation » (Holling, 1973, p. 7), et de la théorie de l'« enchaînement historique » qui suppose que certaines relations historiques perdurent, même si elles ne sont plus optimales, car il serait trop coûteux de rompre ces habitudes prises dans le passé (*path dependence*, Martin, Sunley, 2006 ; Pumain, 2006).

---

[1] L'Europe centrale et orientale que nous retenons est composée des huit pays anciennement communistes qui n'ont pas été membres de l'URSS et qui font partie de l'Union européenne aujourd'hui (Bulgarie, Croatie, Hongrie, Pologne, Roumanie, Slovaquie, Slovénie, Tchéquie).

**Méthodologie et source de données**

Nous analysons le poids de ces relations du passé, selon des périodes distinctes, avant et après la chute du Mur de Berlin selon trois entrées principales pour lesquelles nous avons disposé de sources de données : les réseaux du commerce mondial, du transport aérien international et des firmes transnationales. Ces derniers réseaux sont emblématiques de la mondialisation et représentatifs de l'ouverture économique de l'Europe centrale et orientale après 1989. En effet, les investissements directs à l'étranger ont été particulièrement déterminants dans l'intégration de l'espace centre-est européen dans l'économie européenne et mondiale dans les années 1990 (Krifa, Vermeire, 1998 ; Louis, Lepape, 2004).

*CHELEM* (Comptes Harmonisés sur les Échanges et L'Économie Mondiale), élaborée par le CEPII (Centre d'études prospectives et d'informations internationales) à Paris, recense des données relatives au commerce international. Elle a permis de couvrir, pour chaque année, la totalité des flux d'échanges depuis 1967 jusqu'à 2012, détaillés par pays exportateur, pays importateur et catégories de produits échangés[2]. *CHELEM* est construite et réactualisée par le CEPII, en majeure partie à partir de la base *COMTRADE* des Nations Unies (De Saint Vaultry, 2008).

*OFOD* (On-Flight Origin And Destination), de l'OACI (Organisation de l'Aviation Civile Internationale) de Montréal, a aussi été explorée[3]. Son extraction permet de disposer du nombre de sièges offerts, chaque année, sur les connexions aériennes entre les villes centre-est européennes et d'autres villes entre 1989 et 2013. Les connexions partant des villes du monde entier vers celles d'Europe centre-orientale y sont également renseignées[4].

La base de données *ORBIS* produite par le Bureau Van Dijk (BVD) constitue une base mondiale relative aux liens capitalistiques reliant les entreprises mères et filles. L'extraction d'*ORBIS*[5], recouvre l'année 2013[6]. Elle recense toutes les firmes, situées en dehors de l'Europe centrale et orientale et possédant une partie du capital des firmes en Europe centrale et orientale dans tout type de secteurs, mais aussi les firmes centre-est européennes contrôlant le capital d'autres firmes situées à l'extérieur de cet espace. L'originalité d'*ORBIS* est de rendre accessibles des informations portant à la fois sur les firmes possédantes et sur les firmes possédées, au niveau des villes des huit pays d'Europe centrale et orientale étudiés.

Dans la suite du travail, pour évaluer l'évolution du rôle de la distance dans la mise en place des échanges commerciaux, nous avons modélisé les flux selon un modèle gravitaire[7] (Encadré 1). En vertu de ce modèle, « les échanges entre deux régions ou deux villes seront

---

[2] Elle a été acquise par l'intermédiaire du groupe de travail « Intégrations régionales » du Labex Dynamite coordonné par Y. Richard, que nous remercions.
[3] Elle a été acquise par l'intermédiaire de l'ERC GeoDiverCity coordonnée par D. Pumain, que nous remercions.
[4] L'OACI agrège dans cette base de données tous les aéroports d'une ville concernée. Par exemple à Paris, tous les aéroports sont agrégés (CDG, ORLY).
[5] Cette extraction été mise à disposition pour notre travail dans le cadre d'une convention de coopération entre l'équipe de C. Rozenblat à l'Université de Lausanne et celle de l'ERC GeoDiverCity coordonée par D. Pumain. Nous remercions C. Rozenblat.
[6] En réalité il s'agit de données collectées un à deux ans avant cette date pour certaines firmes.
[7] Au vue de la taille de l'échantillon (n=41 en 1989, n=64 en 1999 et n=145 en 2013), un modèle plus complexe à double contrainte de Wilson n'a pas été nécessaire ici

d'autant plus importants que le poids des villes ou des régions est grand et d'autant plus faibles qu'elles seront éloignées » (Pumain, 2004, p. 1).

**Encadré 1** Modèle gravitaire – calcul des flux théoriques[8]

$$Fij = k * \frac{MiMj}{Dij^a}$$

Fij est le flux théorique entre le pays d'origine i et le pays de destination j k est la constante indiquant le niveau global de mobilité qui permet de fixer celui-ci par rapport aux flux réels $k = \frac{\Sigma\ Flux\ réels}{\Sigma\ MiMj}$

Mi et Mj sont les poids des deux pays, ici le PIB

Dij est la distance kilométrique à vol d'oiseau entre les deux pays

Les paramètres des variables du modèle $(\beta, \gamma, \delta)$ ont été ensuite estimé par l'intermédiaire d'une régression linéaire généralisée de Poisson (D'Aubigny et al., 2000). La régression se résume par la formule suivante, où la variable à expliquer est le flux observé et les variables explicatives sont les PIB (*Mi* et *Mj*) et la distance (*Dij*) et le terme d'erreur $u_{ij}$ suit une distribution de Poisson :

$$\log Fij = \alpha \log k + \beta \log Mi + \gamma \log Mj + a\ \delta \log Dij\ + u_{ij}$$

La performance de ce type d'estimation été largement établie par rapport à une régression simple (Flowerdew, Aitkin, 1982). La pertinence de cette approche statistique a été également vérifiée pour les modèles gravitaires en géographie (D'Aubigny et al., 2000). Nous ne rencontrons par ailleurs pas le problème d'un grand nombre de flux nuls pour lequel l'estimateur devrait être adapté (Martin, Pham, 2015).

**Un renforcement du commerce régional depuis les années 2000**

Les relations commerciales entretenues avec les autres pays centre-est européens ont perdu de l'importance dans les années 1990 au profit de relations dirigées vers l'Allemagne et l'Union européenne. Ils ont pu toutefois se réinscrire dans une continuité avec le passé dans les années 2000.

La part des échanges vers les autres pays d'Europe centrale et orientale a chuté au cours des années 1980, avant de connaître un quasi-triplement à partir des années 1990 (Figure 1). Cela s'explique probablement par la décision des chefs d'Etats de la Hongrie, de la Pologne et de la Tchécoslovaquie, de mettre en œuvre une politique de coopération économique en février

---

[8] La distance kilométrique entre les villes a été calculée sous ArcMap grâce à l'outil de génération de table de proximité. Une valeur par défaut *a*=2 a été choisie sans entreprendre son ajustement.

1991 à Višegrad[9] (Lepesant, 2011). En 1993, cette dernière a abouti à la signature d'un accord de libre-échange régional : le Central-European Free Trade Agreement (CEFTA).

**Figure 1** Flux d'exportations à l'intérieur de l'Europe centrale et orientale en 1967, 1992 et 2012, exprimés en millions de dollars constants en 2012

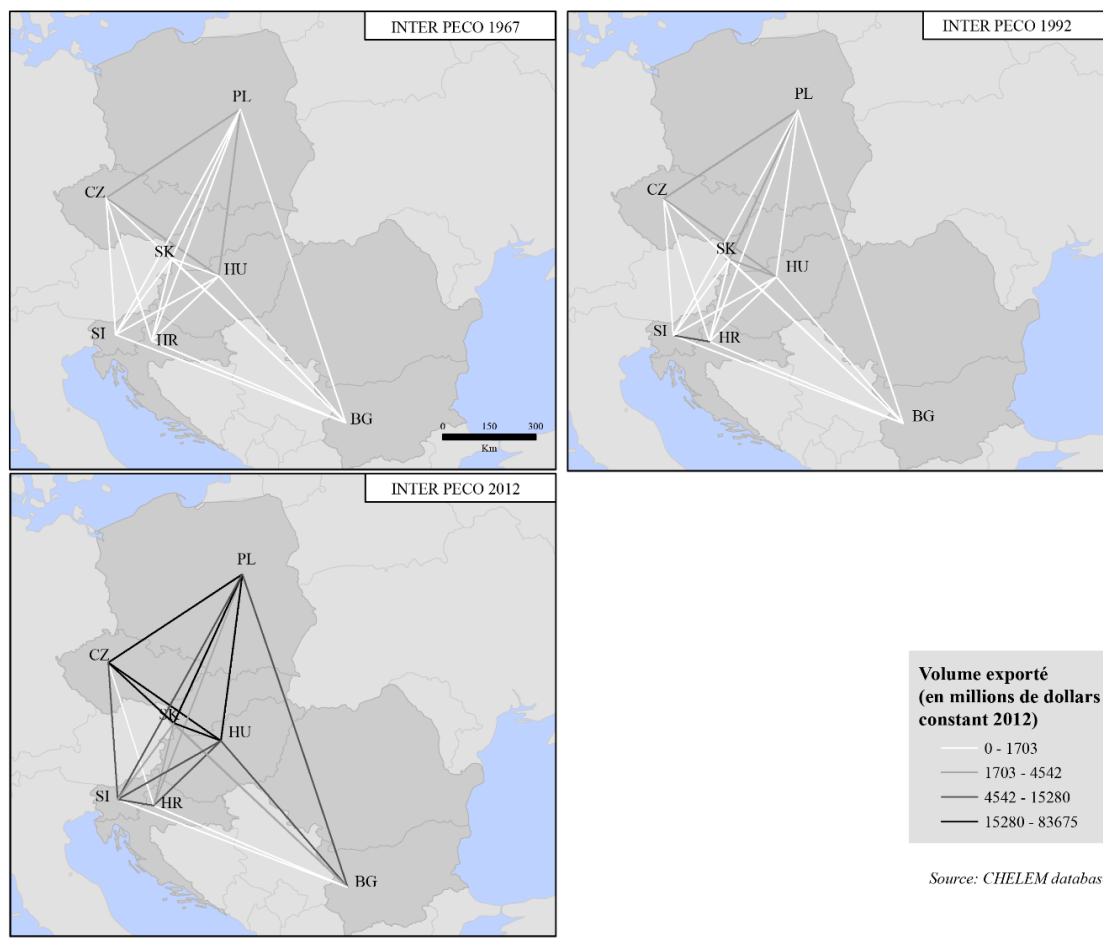

Source: CHELEM database

Le groupe dit de Višegrad s'est élargi à d'autres membres, comme la Slovénie en 1996, puis la Roumanie en 1997. Son succès, atténué par les efforts consentis pour l'intégration à l'Union européenne, n'est pas lié à l'implication stricte des pays membres à coopérer entre eux, mais plutôt à des incitations encourageant des Etats traditionnellement partenaires à créer des « mini-marchés régionaux » (Lepesant, 2011, p. 158). Cette logique a été un moteur pour l'émergence d'un espace balkanique de plus en plus intégré à partir de 2000. Après le départ des pays d'Europe centre-orientale du CEFTA pour rejoindre l'Union européenne en 2004, les pays de l'ex-Yougoslavie, ainsi que la Moldavie ont rejoint cet accord de libre-échange. Cette dynamique est visible à travers l'augmentation du volume exporté, jusqu'en 2012, vers la Slovénie et la Croatie (Figure 1).

Cependant, il convient de noter que les principaux échanges commerciaux entre pays centre-est européens se sont opérés entre la Pologne, la Tchéquie, la Hongrie et la Slovaquie sur toute la période (Figure 1). La principale augmentation du volume échangé intervient en 2012. La majorité des exportations sont alors issues de la Tchéquie, puis de la Pologne. À l'inverse la part des exportations issues de la Bulgarie, Croatie et Slovénie a été très faible. Il semblerait qu'il existe ainsi une région plus intégrée en Europe centre-orientale située dans le Nord-Ouest.

---

[9] Višegrad est un village en Hongrie, où a eu lieu cette décision. Par extension on parle jusqu'à aujourd'hui du groupe de Višegrad (triangle de Višegrad ou V4) pour parler de la Hongrie, la Pologne, la Slovaquie et la Tchéquie, à l'origine de ce groupe.

Si l'on reprend l'hypothèse du modèle gravitaire (voir Encadré 1) deux facteurs principaux semblent déterminer le volume échangé : le poids économique des pays et la distance qui les sépare. Pour vérifier cette hypothèse, nous avons considéré les PIB respectifs de chaque couple de pays centre-est européens échangeant entre eux et la distance qui sépare leurs capitales. Les résultats montrent que le coefficient de la distance ($\delta$) a diminué entre 1967 et 2012, ce qui signifie que le frein de celle-ci a été de plus en plus important pour la mise en place des flux commerciaux entre pays centre-est européens (Tableau 1).

**Tableau 1** Estimation des paramètres des variables de population ($\beta$ et $\gamma$) et de la distance ($\delta$) de la régression et coefficients de détermination en 1967, 1992, 2002 et 2012[10]

|   | 1967 | 1992 | 2002 | 2012 |
|---|---|---|---|---|
| $\beta$ | 0,8 | 0,5 | 0,8 | 1,2 |
| $\gamma$ | 0,8 | 0,5 | 0,8 | 1,1 |
| $\delta$ | 0,4 | -1,7 | -1,1 | -1,8 |
| $R^2$ | 0,6 | 0,6 | 0,7 | 0,8 |

*Source : réalisé par l'auteur à partir de la base de données CHELEM*

De plus l'influence du PIB s'est renforcée en 2012 au vue de l'accroissement des coefficients ($\beta$ et $\gamma$). Enfin la qualité de la relation a augmenté cours du temps – notamment après les années 2000 ($R^2 = 0,8$ en 2012). Les flux sont ainsi devenus plus polarisés sur les pays les plus développés et proches, ce qui explique les importants échanges entre la Pologne, la Tchéquie, la Hongrie et la Slovaquie en 2012 (Figure 1). Cela permet de confirmer un effet de régionalisation en Europe centre-orientale autour de ces quatre pays.

Nous allons vérifier, dans la prochaine section, si cette relative continuité des relations au sein de celui de l'Europe centrale et orientale se remarque également dans le cadre des connexions aériennes au niveau des villes.

**Un trafic aérien en expansion dès les années 2000**

Dans les années 1990, le trafic aérien très compétitif et hiérarchisé au niveau européen et mondial (Matsumoto, 2007) n'en était qu'à ses prémices de développement, en Europe centrale et orientale. Les liaisons aériennes entre villes d'Europe centre-orientale étaient peu nombreuses jusqu'en 1989 (Baker, Prath, 1989). Elles ont ensuite connu un développement spectaculaire jusqu'en 2013 (Figure 2). Le trafic a plus que doublé entre 1989 et 2013. Ces villes se sont retrouvées en deuxième destination du trafic, après les villes de l'espace de l'Union européenne (hors PECO), entre 1990 et 2013.

---

[10] L'intervalle de confiance est de 95%.

**Figure 2** Trafic aérien international entre villes d'Europe centre-orientale en 1989, 1999 et 2013[11]

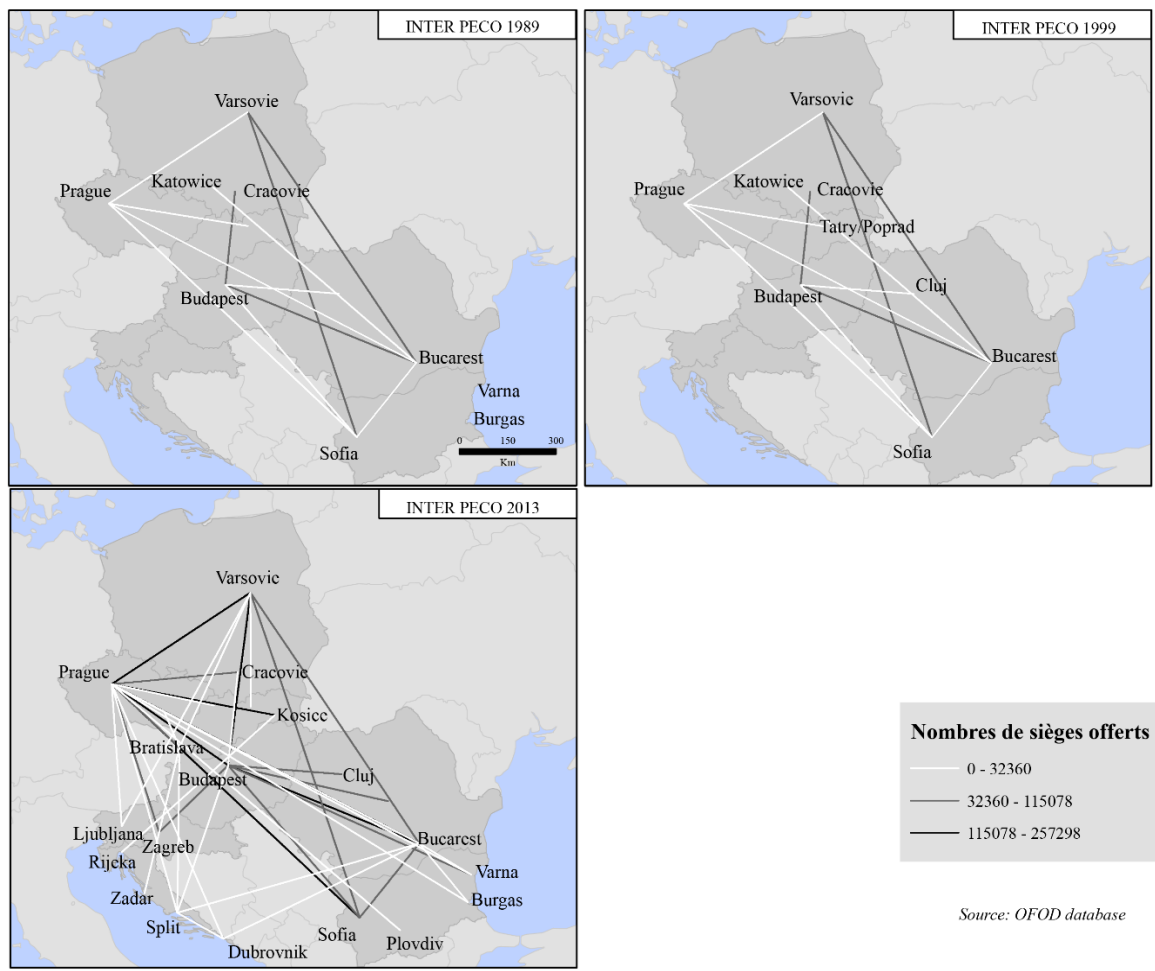

En 1989, les principales connexions se concentraient autour de trois capitales : Budapest, Prague et Varsovie (Figure 2). Il s'agit des premières villes ayant développé des caractéristiques métropolitaines en Europe centrale et orientale (Bauer, Marciszewska, 2001). Dix ans après, en 1999, en pleine crise économique, les flux ont été moins concentrés sur ces villes et se sont renforcés à l'Est de l'espace centre-est européen, comme le montre le nombre élevé de sièges entre Varsovie, Bucarest et Sofia, ainsi qu'entre Budapest et Bucarest (Figure 2). Cette multiplication des flux est liée à l'intense développement du transport aérien en Roumanie et en Bulgarie lors de cette période (Rey, Groza, 2008).

Dans les années 2000, les liaisons entre Prague, Budapest et Varsovie ont généré le plus grand nombre de sièges. Cette concentration peut s'expliquer notamment par la croissance de l'attractivité touristique et économique (Rozenblat et al., 2008). Les autres villes de l'espace centre-est européen se sont davantage connectées que par le passé, et notamment avec les villes du Sud (Figure 2). Il s'agit principalement de Ljubljana, Zagreb, Rijeka, et d'autres destinations croates comme Zadar, Split et Dubrovnik. Ce changement a essentiellement été impulsé par le développement du tourisme sur la côte méditerranéenne (Baulant et al., 2013). Les connexions avec Bratislava ou Kosice, d'abord inexistantes à cause de la centralisation du trafic aérien à Prague ou Vienne, se sont développées à partir des années 2010 (Bauer, Marciszewska, 2001).

---

[11] Le choix des dates se justifie par la première et dernière disponible dans *OFOD*. Celle de 1999 correspond à l'année de la crise économique mondiale.

Ainsi, le développement du marché aérien en Europe centre-orientale a eu tendance à stimuler des interactions nouvelles avec des villes de l'espace centre-est européen autres que les capitales. Nous vérifierons enfin si ces dynamiques se retrouvent également dans le cadre des réseaux de firmes transnationales.

**Une régionalisation de l'espace centre-est européen au sein des réseaux de firmes transnationales en 2013**

En 2005, la première firme polonaise de raffinerie de pétrole PKN ORLEN a racheté UNIPETROL, troisième firme tchèque en termes de capitalisation boursière du pays. Cela constitue le signe d'une interaction économique de taille entre firmes situées dans l'espace centre-est européen. Cette section sera l'occasion d'analyser ce type de relations.

En termes de liens de contrôle entre villes centre-est européennes, les relations apparaissent très asymétriques. Les liens de contrôle les plus forts relient Prague, Varsovie et Budapest et révèlent un clair effet de régionalisation dans l'Europe du centre-ouest (Figure 3).

**Figure 3** Liens de contrôle de capital entre firmes centre-est européennes en 2013 agrégés au niveau des villes

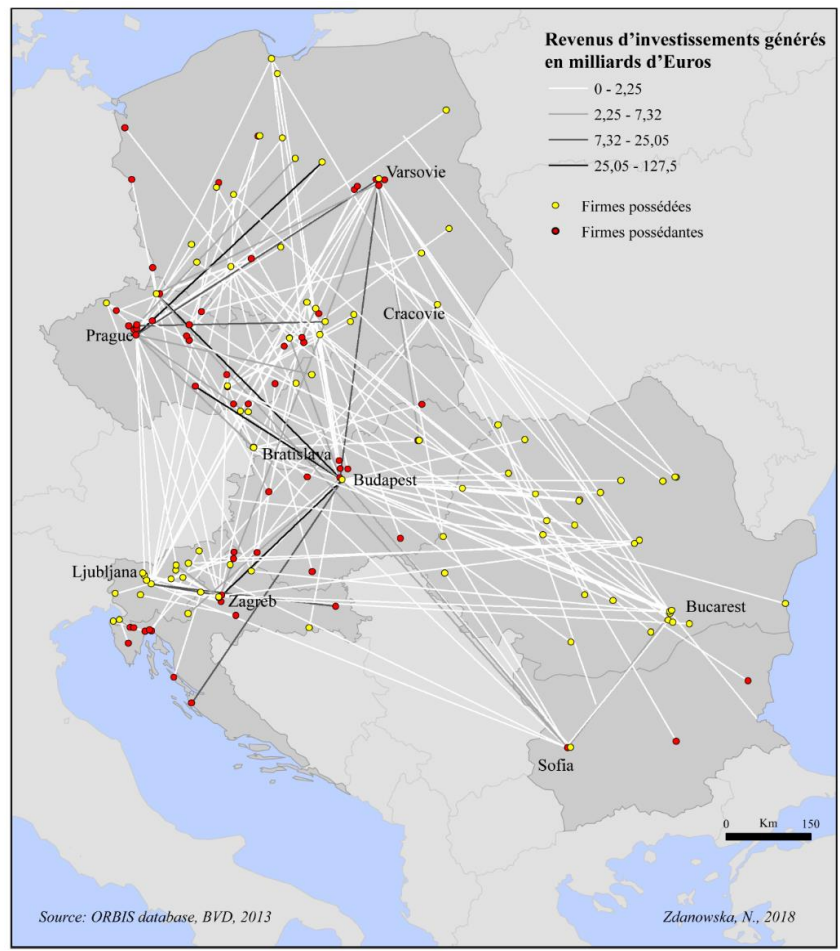

Les contextes historiques et les relations privilégiées durant les hégémonies autrichienne, prussienne et ottomane aux XVIIIe et XIXe siècles dans l'espace centre-est européen, ainsi que pendant la période socialiste, sont certainement encore aujourd'hui responsables de l'orientation des liens de contrôle du capital parmi les pays d'Europe centre-orientale (Lepesant, 2011 ; Zdanowska, 2015). Les firmes roumaines par exemple, sont pour la plupart possédées par les firmes hongroises en 2103, ce qui a généré 235 millions d'euros de revenus d'investissements directs étrangers. Les territoires actuels des deux pays, en particulier la Transylvanie, possédaient de nombreux points communs à l'époque de l'Empire austro-hongrois. De même, les firmes slovaques possèdent une grande partie des capitaux des firmes tchèques (390 millions d'euros) en Europe centre-orientale en 2013. L'inverse se vérifie également (383 millions d'euros pour les firmes tchèques en Slovaquie). Ceci reflète les relations privilégiées qu'elles ont développées au cours de la période tchécoslovaque. Les cas de la Slovénie et de la Croatie présentent des similitudes : les capitaux possédés par la Slovénie dans les entreprises croates représentent 2 milliards d'euros, tandis que les capitaux inverses représentent 128 millions d'euros. Là encore, l'histoire commune de ces deux pays au sein de l'ex-Yougoslavie en constitue probablement la cause. Tous ces exemples témoignent de la continuité de ces relations mises en place auparavant, sous des régimes politiques et économiques aujourd'hui disparus (voir aussi Lepesant, 2011).

Ces asymétries de liens de contrôle s'expliquent, également, par les différences de niveaux de développement entre chaque pays centre-est européen (Lux, 2010). Il semblerait, en effet, que la Roumanie ou la Bulgarie ne disposent pas des mêmes aptitudes à investir sur des marchés étrangers, y compris au niveau centre-est européen, que la Hongrie, la Pologne ou la Tchéquie. Toutes ces observations confirment l'existence d'un effet de régionalisation au sein de l'Europe centrale et orientale.

**Conclusion**

Cet article a permis de mettre en évidence une relative continuité des relations des villes centre-est européennes depuis l'ère communiste, malgré un ralentissement général dans les années 1990 suite aux changements politiques et économiques. Les traces du passé ont donc bien perduré jusqu'à aujourd'hui.

Les échanges commerciaux entre pays centre-est européens, se sont, quant à eux, renforcés au cours du temps après la chute du Mur et dans les années 2000, grâce à des accords politiques favorables comme le CEFTA du groupe de Višegrad, ou par l'intégration au marché commun de l'Union européenne. Cela illustre une reproduction des traces du passé dans un contexte libéral, liée notamment aux habitudes d'échanges prises et au surcoût qui découlerait d'un changement (Hall, Taylor, 1997). Le rôle important joué par la proximité géographique et la richesse des pays sur les échanges centre-est européens a aussi été relevé. Ceci a permis de confirmer l'hypothèse d'un effet de régionalisation dans l'espace centre-est européen autours de la Hongrie, la Pologne, la Slovaquie et la Pologne.

L'orientation des connexions aériennes, depuis 1989, s'inscrit dans la même logique. Bien que le transport aérien ait été en plein essor au début des années 1990, il a connu un développement et une diversification spectaculaires jusqu'en 2013. Toutefois, par son caractère très hiérarchique et concurrentiel, l'offre aérienne s'est concentrée sur les plus grandes villes au cours du temps dans le cadre des connexions au sein de l'espace centre-est européen. En effet, à cause des coûts d'installation et de maintenance très élevés, le transport aérien est parfois moins rentable sur de courtes distances et dans des zones périphériques que, par exemple, l'autobus repris par des compagnies privées (Menes, 2000).

En termes de revenus d'IDE générés, les liens capitalistiques entre firmes centre-est européennes se sont avérés extrêmement asymétriques et concentrés sur la partie centrale de l'espace centre-est européen, autour de la Pologne, la Tchéquie et de la Hongrie, confirmant de nouveau un effet de régionalisation dans la Nord-Ouest.

**Bibliographie**